\documentclass[preprint,showpacs]{revtex4-1}

\usepackage{graphicx}
\usepackage{dcolumn}
\usepackage{mathrsfs}

\usepackage{slashed}
\usepackage{amsmath,epsfig}
\usepackage{amsfonts}
\usepackage{amssymb}
\usepackage{graphicx}
\usepackage{dcolumn}
\usepackage{bm}
\def\beq{\begin{equation}}
\def\eeq{\end{equation}}
\def\bea{\begin{eqnarray}}
\def\eea{\end{eqnarray}}
\def\nn{\nonumber}
\def\bge{\begin{equation}}
\def\ene{\end{equation}}
\def\bg{\begin{eqnarray}}
\def\en{\end{eqnarray}}

\def\q{{\bf q}}

\def\q{{\bf q}}

\def\cbar{\bar{c}}
\def\qbar{\bar{q}}

\def\D0bar{\overline{D^0}}

\begin{document}

\vspace*{-2em}
\hfill ADP-10-12/T708, JLAB-THY-10-1188
\vspace{1em}

\title{$J/\Psi$ mass shift in nuclear matter}
\author{
G. Krein$^1$\footnote{gkrein@ift.unesp.br},
A.W. Thomas$^2$\footnote{anthony.thomas@adelaide.edu.au}
and K. Tsushima$^3$\footnote{tsushima@jlab.org}}
\affiliation{$^1$Instituto de F\'{\i}sica Te\'orica, Universidade Estadual Paulista \\
Rua Dr. Bento Teobaldo Ferraz, 271 - Bloco II, S\~ao Paulo, SP, Brazil \\
$^2$CSSM, School of Chemistry and Physics,
University of Adelaide, Adelaide SA 5005, Australia\\
$^3$EBAC in Theory Center, Jefferson Lab, 12000 Jefferson Ave. Newport News,
VA 23606, USA
}

\begin{abstract}
The $J/\Psi$ mass shift in cold nuclear matter is computed using an effective
Lagrangian approach. The mass shift is computed by evaluating $D$ and $D^*$
meson loop contributions to the $J/\Psi$ self-energy employing medium-modified
meson masses. The modification of the $D$ and $D^*$ masses in nuclear matter is
obtained using the quark-meson coupling model. The loop integrals are regularized
with dipole form factors and the sensitivity of the results to the values of
form-factor cutoff masses is investigated. The $J/\Psi$ mass shift arising from
the modification of the
$D$ and $D^*$ loops at normal nuclear matter density is found to range from
$-16$~MeV to $-24$~MeV under a wide variation of values of the cutoff masses.
Experimental perspectives for the formation of a bound state of $J/\Psi$ to a
nucleus are investigated.
\end{abstract}

\pacs{21.65.Jk, 21.85.+d, 24.85.+p, 14.40.Pq, 14.40.Lb}

\maketitle

\section{Introduction}
\label{sec:intro}

A new era of nuclear matter research is envisaged with the $12$~GeV
upgrade of the CEBAF accelerator at the Jefferson Lab in the USA and
with the construction of the FAIR facility in Germany. These new
facilities will have the exciting potential of implanting low-momentum
charmonia and charmed hadrons in an atomic nucleus, like the $J/\Psi$ and
$\psi$ mesons and heavy-light charmed mesons such as $D$ and $D^*$. While
at JLab charmed hadrons will be produced by scattering electrons off
nuclei, at FAIR they will be produced by the annihilation of antiprotons
on nuclei. There are several reasons for the excitement, one of the
main ones being the opportunity of studying the poorly understood
low-energy excitations of gluon degrees of freedom. An example where
these excitations play an important role is the propagation of charmonia
in matter. Since a charmonium state does not have quarks in common with
the nuclear medium, its interactions with the medium necessarily involve
the intervention of gluons. Basic interaction mechanisms discussed in
the literature have been the excitation of QCD van der Waals forces
arising from the exchange of two or more gluons between color-singlet
states~\cite{Peskin:1979va,Brodsky:1989jd}, and the excitation of
charmed hadronic intermediate states with light quarks created from the
vacuum~\cite{Brodsky:1997gh,Ko:2000jx}.

Another interesting challenge is
to study the properties of charmed $D$ and $D^*$ mesons in medium. The chiral
properties of the light quarks that compose these mesons are much more sensitive
to the nuclear medium than their companion, heavier charm quarks and
therefore they offer the unique opportunity of studying phenomena like
the partial restoration of chiral symmetry in nuclear matter.
Motivated by such considerations, some very interesting phenomena
involving these mesons have been
predicted. Amongst these we mention the possible formation
of $D(\bar{D})$ meson-nuclear
bound states~\cite{Tsushima:1998ru}, enhanced dissociation of $J/\Psi$ meson
in nuclear matter (heavy nuclei)~\cite{Sibirtsev:1999jr},
and enhancement of the $D$ and $\bar{D}$ meson production in antiproton-nucleus
collisions~\cite{Sibirtsev:1999js}. Ref.~\cite{Voloshin:2007dx} presents a recent
review of the properties of charmonium states and compiles a
fairly complete list of references on theoretical studies concerning a
great variety of physics issues related to these states. On the experimental
side, one of the major challenges is to find appropriate kinematical conditions
to produce these hadrons essentially at rest, or with small momentum
relative to the nucleus, as effects of the nuclear medium are driven
by low energy interactions.

The original suggestion~\cite{Brodsky:1989jd} was that
QCD van der Waals forces arising from multiple gluon exchange
would be capable of binding a charmonium state
by as much as $400$~MeV in an $A=9$
nucleus. The estimate was based on a variational calculation using a
phenomenological ansatz for the charmonium-nucleus potential in the
form of a Yukawa potential. Along the same lines but taking into account
the distribution of nucleons in the nucleus by folding the
charmonium-nucleon Yukawa potential with the nuclear density
distribution, Ref.~\cite{Wasson:1991fb} found a maximum of $30$~MeV
binding energy in a large nucleus. A somewhat more QCD-oriented
estimate was made in Ref.~\cite{Luke:1992tm}. Using a lowest-order
multipole expansion for the coupling of multiple gluons to a small-size
charmonium bound state~\cite{Peskin:1979va}, it is possible to show on
the basis of the operator product expansion that the mass shift of
charmonium in nuclear matter is given, in the limit of infinitely heavy charm quark
mass, by an expression similar to the usual second-order Stark effect in
atomic physics, which depends on the chromo-electric polarizability of the
nucleon. Using an estimate~\cite{Peskin:1979va} for the value of this
polarizability, the authors of Ref.~\cite{Luke:1992tm} obtained a $10$~MeV
binding for $J/\Psi$ in nuclear matter. On the other hand, for the excited
charmonium states, a much larger binding energy was obtained, e.g. $700$~MeV
for the excited charmonium state $\psi'(2S)$, an admittedly untrustworthy
number. Following this same procedure, but keeping the charm quark mass
finite and using realistic charmonium bound-state wave-functions,
Ref.~\cite{Ko:2000jx} found $8$~MeV binding energy for $J/\Psi$ in
nuclear matter, but still over $100$~MeV binding for the charmonium excited
states. While an increase in the QCD Stark effect is expected for
excited states (because of their larger size), the extreme
values for the binding energies for these states found in the literature
are widely considered to be unrealistic. The source for such an overestimate is
attributed to the breakdown of the multipole expansion for the
larger-sized charmonium states.

There are some other studies on charmonium interactions with ordinary
hadrons and nuclear matter, in particular involving the $J/\Psi$ meson.
QCD sum rules studies estimated a $J/\Psi$ mass decrease in nuclear matter
ranging from $4$ to $7$ MeV~\cite{Klingl:1998sr,Hayashigaki:1998ey,Kim:2000kj},
while an estimate based on color polarizability~\cite{Sibirtsev:2005ex}
gave larger than $21$ MeV. In addition, there are studies of the charmonium-nucleon
interaction and of $J/\Psi$ dissociation cross sections
based on a one-boson exchange
model~\cite{Sibirtsev:2000aw},
effective Lagrangians~\cite{Liu:2001ce,Oh:2007ej} and
the quark-model~\cite{Hilbert:2007hc}.
In Ref.~\cite{Yokokawa:2006td} the charmonium-hadron
interaction was studied in lattice QCD.

\begin{figure}[htb]
  \includegraphics[height=0.15\textheight]{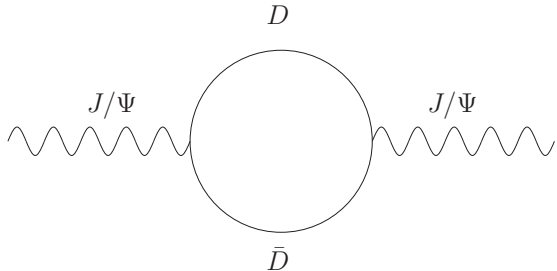}
  \caption{$DD$-loop contribution to the $J/\Psi$ self-energy. We include
  also  $DD^*$ and $D^*D^*$ contributions.}
  \label{fig:loop}
\end{figure}

A first estimate for the mass shifts of charmonium states (we denote
charmonium states generically by $\psi$) in nuclear medium arising from the
excitation of a pair of $D$ and $D^*$ mesons -- see Fig.~\ref{fig:loop}
-- was performed in Ref.~\cite{Ko:2000jx}. Employing a gauged effective
Lagrangian for the coupling of $D$ mesons to the charmonia, the mass shifts
were found to be positive for $J/\Psi$ and $\psi(3770)$, and negative for
$\psi(3660)$ at normal nuclear matter density $\rho_0$. These results were
obtained for density-dependent $D$ and $\bar D$ masses that decrease linearly
with density, such that at $\rho_0$ they are shifted by $50$~MeV. The loop
integral in the self-energy (Fig.~\ref{fig:loop}) is divergent and
was regularized using form-factors derived from the $^3P_0$ decay
model with quark-model wave functions for $\psi$ and $D$. The positive
mass shift is at first sight puzzling, since even with a $50$~MeV
reduction of the $D$ masses, the intermediate state is still
above threshold for the decay of $J/\Psi$ into a $D\bar D$ pair and
so a second-order contribution should be negative. As we shall explain
below, this was not realized in the calculation of Ref.~\cite{Ko:2000jx}
because of the interplay
of the form factor used and the gauged nature of the interaction.
\vspace{0.5cm}

In the present paper we reanalyze the mass shift of $J/\Psi$ in terms
of the excitation of intermediate charmed mesons using
effective Lagrangians. In addition to the $D\bar D$ loops, we also include
$D \bar D^*$, $D^* \bar D$ and $D^*{\bar D}^*$ loops. The medium dependence
of the $D$ and $D^*$ masses is included by an explicit calculation using
the quark-meson coupling (QMC) model~\cite{Guichon:1987jp}. The QMC is a
quark-based model for nuclear structure which has been very successful
in describing nuclear matter saturation properties and has been used to
predict a great variety of changes of hadron properties in nuclear medium.
A~review of the basic ingredients of the model and a summary of results and
predictions can be found in Ref.~\cite{Saito:2005rv}.

The paper is organized as follows. In the next Section we present the effective
Lagrangians used to calculate the $J/\Psi$ self-energy and give explicit
expressions for the contributions of the different intermediate states. In
Section~\ref{sec:qmc} we briefly review the QMC description of the $D$ and $D^*$
mesons in nuclear matter and present numerical results for the density dependence
of the $D$ and $D^*$ masses. A full set of numerical results for the density
dependence of the $J/\Psi$ self-energy is
presented in Section~\ref{sec:num}. We show results
for the separate contributions of
the $D \bar D^*$, $D^* \bar D$ and $D^*{\bar D}^*$
loops and also investigate the sensitivity of our results to the cutoff masses.
Our conclusions and perspectives for future work are presented in
Section~\ref{sec:concl}.

\section{Effective Lagrangians and $J/\Psi$ self-energy}
\label{sec:lag}

We use the following phenomenological Lagrangian densities for the vertices $J/\Psi$-$D$
and $J/\Psi$-$D^*$ (in the following we denote by $\psi$ the field representing
$J/\Psi$):
\begin{widetext}
\bea
{\mathcal L}_{\psi D D} &=& i g_{\psi D D} \, \psi^\mu
\left[\bar D \left(\partial_\mu D\right) -
\left(\partial_\mu \bar D\right) D \right] ,
\label{LpsiDDbar} \\
{\cal L}_{\psi D D^*} &=& \frac{ g_{\psi D D^*}}{m_\psi} \,
\varepsilon_{\alpha\beta\mu\nu} \left(\partial^\alpha \psi^\beta\right)
\Bigl[\left(\partial^\mu \bar{D}^{*\nu}\right) D +
\bar D \left(\partial^\mu D^{*\nu}\right)  \Bigr] ,
\label{LpsiDD*} \\
{\cal L}_{\psi D^* D^*} &=& i g_{\psi D^* D^*} \, \bigl\{ \psi^\mu
\left[\left(\partial_\mu \bar{D}^{*\nu}\right) D^*_\nu
- {\bar D}^{*\nu}\left(\partial_\mu D^*_\nu\right) \right]
\nn\\
& &\hspace{3em}+ \left[ \left(\partial_\mu \psi^\nu\right) \bar{D}^*_\nu
- \psi^\nu \left(\partial_\mu {\bar D}^*_\nu\right)
\right] D^{*\mu}
+ \, \bar{D}^{*\mu} \left[\psi^\nu \left(\partial_\mu D^*_\nu\right)
- \left(\partial_\mu \psi^\nu\right) D^*_\nu \right]
\bigr\} .
\label{LpsiD*D*}
\eea
\end{widetext}
Our convention for the $D$-meson-field isospin doublets is
\bea
\bar D = ({\bar D}^0 \;\;\; D^{-} ),
\hspace{1.0cm}
{D = \left(\begin{array}{c}
  D^0 \\
  D^+
\end{array}\right)} .
\label{doublets}
\eea
We note that these Lagrangians are an $SU(4)$ extension of light-flavor
chiral-symmetric Lagrangians of pseudoscalar and vector mesons. In the light
flavor sector, they have been motivated by a local gauge symmetry principle,
treating vector mesons either as massive gauge bosons or as dynamically
generated gauge bosons. In the first case, there appear contact interactions
involving two pseudoscalar and two vector mesons. When extended to the charm sector,
in Eq.~(\ref{LpsiDDbar}) for instance, there is an additional term of
the form $2 g^2_{\psi D  D} \psi^\mu\psi_\mu \bar D D$. In view of the fact that $SU(4)$
flavor symmetry is strongly broken in nature, and in order to stay as close as possible
to phenomenology, we use experimental values for the charmed mesons masses and use the
empirically known meson coupling constants. For these reasons we choose not to use gauged
Lagrangians -- a similar attitude was followed in Ref.~\cite{Lin:1999ve} in a study of
hadronic scattering of charmed mesons. However, in order to compare results with
Ref.~\cite{Ko:2000jx} and assess the impact of a contact term of the form $\psi \psi D D$,
we will also present results for the $J/\Psi$ mass shift including such a term.

We are interested in the difference of the in-medium, $m^*_\psi$, and vacuum,
$m_\psi$, masses of $J/\Psi$,
\beq
\Delta m = m^*_\psi - m_\psi,
\label{Delta-m}
\eeq
with the masses obtained from
\beq
m^2_\psi = (m^0_\psi)^2 + \Sigma(k^2 = m^2_\psi)\, .
\label{m-psi}
\eeq
Here $m^0_\psi$ is the bare mass and $\Sigma(k^2)$ is the total $J/\Psi$
self-energy obtained from the sum of the contributions from the $DD$, $DD^*$ and
$D^*D^*$ loops. The in-medium mass, $m^*_\psi$, is obtained likewise, with the
self-energy calculated with medium-modified $D$ and $D^*$ meson masses.

We take the averaged, equal masses for the neutral and charged $D$ mesons, i.e.
$m_{D^0} = m_{D^{\pm}}$ and $m_{D^{*0}} = m_{D^{*\pm}}$. Averaging over the
three polarizations of $J/\Psi$, one can write each of the loop contributions
to the $J/\Psi$ self-energy $\Sigma_l$, $l=DD,DD^*,D^*D^*$, as
\beq
\Sigma_l(m^2_\psi) = - \frac{ g^2_{\psi \, l}}{3\pi^2}
\int^\infty_0 dq \, \q^2 \, F_l(\q^2) \, K_l(\q^2),
\label{Sigma-l}
\eeq
where $F_l(\q^2)$ is the product of vertex form-factors (to be discussed later) and the
$K_l(\q)$ for each loop contribution are given by
\bea
K_{DD}(\q^2) &=& \frac{\q^2}{\omega_D} \left( \frac{\q^2}{\omega^2_D
- m^2_\psi/4} - \xi \right) , \label{KDD} \\
K_{DD^*}(\q^2) &=& \frac{\q^2 \, \overline{\omega}_D}
{\omega_D \omega_{D^*}}\frac{1}{\overline{\omega}^2_D
- m^2_\psi/4}, \label{KDD*} \\
K_{D^*D^*}(\q^2) &=& \frac{1}{4 m_\psi \omega_D }
\Biggl[\frac{{\cal A}(q^0=\omega_{D^*})}{\omega_{D^*} - m_\psi/2}
- \, \frac{{\cal A}(q^0=\omega_{D^*}+m_\psi)}{\omega_{D^*} + m_\psi/2} \Biggr],
\label{SigmaD*D*}
\eea
where $\omega_D = (\q^2+m^2_D)^{1/2}$, $\omega_{D^*} = (\q^2+m^2_{D^*})^{1/2}$,
$\overline\omega_D = (\omega_D + \omega_{D^*})/2$,
$\xi = 0$ for the non-gauged Lagrangian of Eq.~(\ref{LpsiDDbar}) and $\xi = 1$
for the gauged Lagrangian of Ref.~\cite{Ko:2000jx}, and
\beq
{\cal A}(q) = \sum^4_{i=1} {\cal A}_i(q),
\eeq
with
\bea
{\cal A}_1(q) &=& - 4 q^2 \left\{  4 - \frac{q^2+(q-k)^2}{m^2_{D^*}}
+ \frac{[q\cdot(q-k)]^2}{m^4_{D^*}}\right\},
\label{A1} \\
{\cal A}_2(q) &=& 8 \left[q^2 - \frac{q\cdot(q-k)}{m^2_{D^*}}\right]
\left[2 + \frac{(q^0)^2}{m^2_{D^*}}\right],
\label{A2} \\
{\cal A}_3(q) &=& 8 \left(2q^0 - m_\psi\right)
\Biggl\{q^0 - \left(2q^0 - m_\psi\right)\frac{q^2+q\cdot(q-k)}{m^2_{D^*}}
\, + q^0 \frac{[q\cdot(q-k)]^2}{m^4_{D^*}}\Biggr\} ,
\label{A3} \\
{\cal A}_4(q) &=& -8 \left[q^0 - (q^0-m_\psi)\frac{q\cdot(q-k)}{m^2_{D^*}}\right]
\, \left[ (q^0 - m_\psi) - q^0 \frac{q\cdot(q-k)}{m^2_{D^*}}\right] .
\label{A4}
\eea
In these last expressions, $q$ and $k$ are four-vectors given by
$q=(q^0,\q)$ and $k = (m_\psi,0)$.

\section{Quark-meson coupling model and $D$ and $D^*$ mesons in matter}
\label{sec:qmc}

In this section we briefly review the QMC description of the $D$ and $D^*$ mesons
in nuclear matter. Notations and explicit expressions
are given in Refs.~\cite{Tsushima:1998ru,Tsushima:2002cc}.

The QMC model was created to provide insight into the structure of nuclear
matter, starting at the quark
level~\cite{Guichon:1987jp,Guichon:1995ue,Saito:2005rv}.
Nucleon internal structure was modeled by the MIT bag,
while the binding was described by the self-consistent couplings of the
confined light quarks ($u,d$) (not $s$ nor heavier quarks) to
the scalar-$\sigma$ and vector-$\omega$ meson fields generated
by the confined light quarks in the other nucleons.
The self-consistent response of the bound light quarks to
the mean $\sigma$ field
leads to a novel saturation mechanism for nuclear matter,
with the enhancement of
the lower components of the valence Dirac light quark wave functions.
The direct interaction between the light quarks and the scalar
$\sigma$ field is a key ingredient of the model, it induces a
nucleon {\it scalar
polarizability}~\cite{Thomas:2004iw,Chanfray:2006nz} and generates
a nonlinear scalar potential (effective nucleon mass), or equivalently
a density-dependent ($\sigma$-field dependent) $\sigma$-nucleon coupling.
The model has opened tremendous opportunities for studies of
the structure of finite nuclei and of hadron properties in a nuclear
medium (nuclei) with a model based on the underlying quark degrees of
freedom~\cite{Saito:2005rv}.

In QMC the Dirac equations for the quarks and antiquarks
in nuclear matter, inside the bags of
$D$ and $D^*$ mesons, ($q = u$ or $d$, and $c$)
neglecting the Coulomb force in nuclear matter, are given by:
\begin{widetext}
\begin{eqnarray}
& &\left[ i \gamma \cdot \partial_x -
(m_q - V^q_\sigma)
\mp \gamma^0
\left( V^q_\omega +
\frac{1}{2} V^q_\rho
\right) \right]
\left( \begin{array}{c} \psi_u(x)  \\
\psi_{\bar{u}}(x) \\ \end{array} \right) = 0,
\label{diracu}\\
& &\left[ i \gamma \cdot \partial_x -
(m_q - V^q_\sigma)
\mp \gamma^0
\left( V^q_\omega -
\frac{1}{2} V^q_\rho
\right) \right]
\left( \begin{array}{c} \psi_d(x)  \\
\psi_{\bar{d}}(x) \\ \end{array} \right) = 0,
\label{diracd}\\
& &\left[ i \gamma \cdot \partial_x - m_c \right]
\psi_c (x)\,\, ({\rm or}\,\, \psi_{\cbar}(x)) = 0.
\label{diracc}
\end{eqnarray}
\end{widetext}
The (constant) mean-field potentials for a light quarks in nuclear matter
are defined by $V^q_\sigma \equiv g^q_\sigma \sigma$,
$V^q_\omega \equiv g^q_\omega \omega$ and
$V^q_\rho \equiv g^q_\rho b$,
with $g^q_\sigma$, $g^q_\omega$ and
$g^q_\rho$ the corresponding quark-meson coupling
constants.

The eigenenergies for the quarks in the $D$ and $D^*$ mesons
in units of $1/R_{D,D^*}^*$ are given by,
\bea
\left( \begin{array}{c}
\epsilon_u \\
\epsilon_{\bar{u}}
\end{array} \right)
&=& \Omega_q^* \pm R^*_{D,D^*} \left(
V^q_\omega
+ \frac{1}{2} V^q_\rho \right),
\\
\left( \begin{array}{c} \epsilon_d \\
\epsilon_{\bar{d}}
\end{array} \right)
&=& \Omega_q^* \pm R^*_{D,D^*} \left(
V^q_\omega
- \frac{1}{2} V^q_\rho \right),
\\
\epsilon_c
&=& \epsilon_{\cbar} =
\Omega_c.
\label{energy}
\eea

Then, the $D$ and $D^*$ meson masses
in a nuclear medium $m^*_{D,D^*}$,
are calculated by
\bea
& &\hspace*{-2em} m_{D,D^*}^* = \sum_{j=q,\bar{q},c,\cbar}
\frac{ n_j\Omega_j^* - z_{D,D^*}}{R_{D,D^*}^*}
+ \frac{4}{3}\pi R_{D,D^*}^{* 3} B,
\\
& &\left. \frac{\partial m_{D,D^*}^*}{\partial R_{D,D^*}}
\right|_{R_{D,D^*} = R_{D,D^*}^*} = 0,
\label{DDsmass}
\eea
where $\Omega_q^*=\Omega_{\bar{q}}^*
=[x_q^2 + (R_{D,D^*}^* m_q^*)^2]^{1/2}\,(q=u,d)$, with
$m_q^*=m_q{-}g^q_\sigma \sigma$,
$\Omega_c^*=\Omega_{\cbar}^*=[x_c^2 + (R_{D,D^*}^* m_c)^2]^{1/2}$,
and $x_{q,c}$ being the bag eigenfrequencies.
$B$ (=(170.0 MeV)$^4$) is the bag constant, $n_q (n_{\qbar})$ and $n_c (n_{\cbar})$
are the lowest mode quark (antiquark)
numbers for the quark flavors $q$ and $c$
in the $D$ and $D^*$ mesons, respectively,
and the $z_{D,D^*}$ parameterize the sum of the
center-of-mass and gluon fluctuation effects and are assumed to be
independent of density.
We choose the values
$(m_q, m_c) = (5, 1300)$ MeV for the current quark masses, and $R_N = 0.8$
fm for the bag radius of the nucleon in free space.
The quark-meson coupling constants, $g^q_\sigma$, $g^q_\omega$
and $g^q_\rho$, are adjusted to fit the nuclear
saturation energy and density of symmetric nuclear matter, and the bulk
symmetry energy~\cite{Saito:2005rv}.
Exactly the same coupling constants, $g^q_\sigma$, $g^q_\omega$ and
$g^q_\rho$, are used for the light quarks in the $D$ and $D^*$ mesons and baryons as in
the nucleon.

\begin{figure}[htb]
\includegraphics[height=95mm,angle=-90]{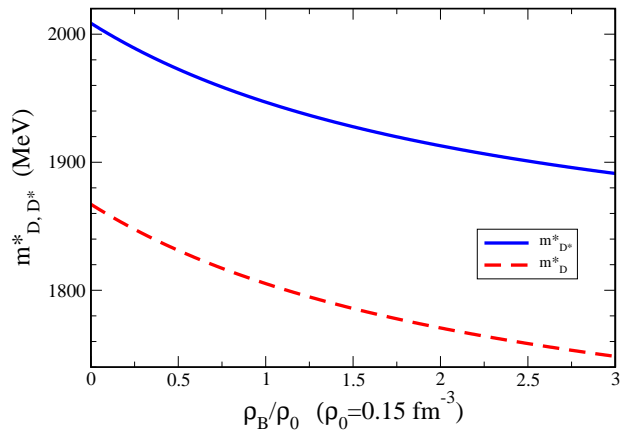}
\caption{
$D$ and $D^*$ meson (scalar) effective masses as a function of
baryon density.
}
\label{fig:DDsmass}
\end{figure}

Because of baryon number conservation, no vector potential should contribute to
the loop integrals. Then, the vector potentials for the $D$ and $D^*$ mesons
should be the same in considering the case of the $D D^*$ mixed loop to cancel out.
However, for the $K^+$ meson case, $g^q_\omega$ associated with the vector potential
had to be scaled $1.96$ times to reproduce an empirically extracted repulsive potential
of about 25 MeV at normal nuclear matter density~\cite{Tsushima:1997df}.
The reason is that $K$-mesons may be regarded as pseudo-Goldstone bosons,
and they are therefore difficult to describe by naive
quark models as is also true for pions.
{}For this reason, in earlier work we explored the possibility of also scaling the
$g^q_\omega$ strength by a factor $1.96$ for the
$D$-mesons~\cite{Tsushima:1998ru,Sibirtsev:1999js}.
In the present case, this possibility is excluded by
baryon number conservation. As a result, the vector potential does not contribute to
the final results. Thus, we may focus on the (scalar) effective masses of
$D$ and $D^*$ mesons. The QMC predictions for the in-medium effective masses of
these mesons are shown in Fig.~\ref{fig:DDsmass}
as a function of nuclear matter density.
The net reductions in the masses of the $D$ and $D^*$
are nearly the same as a function of density,
as dictated by {\it the light quark number counting
rule}~\cite{Tsushima:2002cc}.
%

\section{Numerical results}
\label{sec:num}

Amongst the main ingredients of the present calculation are the phenomenological
form factors needed to regularize the self-energy loop integrals in
Eq.~(\ref{Sigma-l}). Following previous experience with a similar calculation
for the $\rho$ self-energy~\cite{Leinweber:2001ac}, we use a dipole form
for the vertex form factors
\begin{equation}
u_{D,D^*}(\q^2) =  \left(\frac{\Lambda_{D,D^*}^2 + m^2_\psi}
{\Lambda_{D,D^*}^2 + 4\omega^2_{D,D^*}(\q)}\right)^2,
\label{uff}
\end{equation}
so that the $F_l(\q^2)$ in Eq.~(\ref{Sigma-l}) are given by
\bea
&& F_{DD}(\q^2)  = u^2_{D}(\q^2), \\
&& F_{DD^*}(\q^2) = u_D(\q^2) \, u_{D^*}(\q^2), \\
&& F_{D^*D^*}(\q^2) = u^2_{D^*}(\q^2),
\label{ffs}
\eea
where $\Lambda_{D}$ and $\Lambda_{D^*}$ are cutoff masses. Obviously the
main uncertainty here is the value of these cutoff masses. In a simple-minded
picture of the vertices the cutoff masses are related to the extension of
the overlap region of $J/\Psi$ and $D$ mesons at the vertices and therefore
should depend upon the sizes of the wave functions of these mesons. One can
have a rough estimate of $\Lambda_{D}$ and $\Lambda_{D^*}$ by using a quark
model calculation of the form factors. Using a $^3P_0$ model for quark-pair
creation~\cite{LeYaouanc} and Gaussian wave functions for the mesons,
the vertex form factor can be written as~\cite{Ko:2000jx}
\begin{equation}
u_{QM}(\q^2)=e^{-\q^2/4(\beta^2_D+2\beta^2_\psi)},
\label{ff}
\end{equation}
where $\beta_D$ and $\beta_\psi$ are the Gaussian size parameters of the
$D$ and $J/\Psi$ wave functions. Demanding that the $u(\q^2)$ of
Eq.~(\ref{uff}) and $u_{QM}(\q^2)$ have the r.m.s. radii
$\langle r^2 \rangle^{1/2}$, with
\beq
\langle r^2 \rangle = - 6 \, \frac{d \ln u(q^2)}{dq^2}\Bigg|_{\q^2=0},
\eeq
one obtains
\beq
\Lambda^2 = 32(\beta^2_D +2 \beta^2_\psi) - 4m^2_D.
\label{Lambda}
\eeq
Using $m_D = 1867.2$~MeV and for the $\beta$'s the values used in Ref.~\cite{Ko:2000jx},
$\beta_D = 310$~MeV and $\beta_\psi = 520$~MeV, one obtains $\Lambda_D = 2537$~MeV.
Admittedly this is a somewhat rough estimate and it is made solely
to obtain an order of magnitude estimate,
since we do not expect that Gaussian form factors should be very accurate at high
$\q^2$. In view of this and to gauge uncertainties
of our results, we allow the value
of $\Lambda_D$ vary in the range $1000~{\rm MeV} \leq \Lambda_D \leq 3000~{\rm MeV}$.
Moreover, for simplicity we use $\Lambda_D = \Lambda_{D^*}$.

Using $m_{D^*} = 2008.6$ MeV for the average of the vacuum masses of the $D^*$'s,
there remain to be fixed the bare $J/\Psi$ mass $m^0_\psi$ and the coupling constants.
The bare mass is fixed by fitting the physical mass $m_{J/\Psi} = 3096.9$~MeV using
Eq.~(\ref{m-psi}). For the coupling constants we use $g_{\psi DD} = g_{\psi DD^*}
= g_{\psi D^*D^*} = 7.64$, which are obtained by invoking vector-meson-dominance
and use of isospin symmetry~\cite{Lin:1999ad}.

We are now in a position to present the results for the in-medium mass shift
$\Delta m$ of $J/\Psi$, defined in Eq.~(\ref{Delta-m}). We calculate the in-medium
self-energy using the in-medium $D$ meson mass as given by the QMC model
presented in Section~\ref{sec:qmc}. We present results for $\xi = 0$ (no gauge coupling)
and for $\xi = 1$ (with gauge coupling).

\begin{table}[htb]
\begin{tabular}{cc|ccc|c}
\hline
$\;\Lambda_D\;$  & $\;m^*_{J/\Psi}\;$  & $\;DD\;$
& $\;DD^*\;$ & $\;D^*D^*\;$ & $\;\Delta m\;$ \\
\hline
                         $1000$ & $3081$ &  $-3$   & $-2$   & $-11$   & $-16$ \\
                         $1500$ & $3079$ &  $-3.5$ & $-2.5$ & $-12$   & $-18$ \\
                         $2000$ & $3077$ &  $-4$   & $-3$   & $-13$   & $-20$ \\
                         $3000$ & $3072$ &  $-6.5$ & $-5$   & $-12.5$ & $-24$ \\
\hline
\end{tabular}
\caption{In-medium $J/\Psi$ mass $m^*_{J/\Psi}$ and the individual loop
contributions to the mass difference $\Delta m$ at nuclear matter density,
for different values of the cutoff $\Lambda_D$, and using the non-gauged
Lagrangian -- $\xi = 0$ in Eq.~(\ref{KDD}). All quantities are in MeV.
}
\label{tab:xi0}
\end{table}

Initially we present results for $\xi = 0$. In Table~\ref{tab:xi0} we present the in-medium $J/\Psi$
mass $m^*_{J/\Psi}$ and the individual loop contributions to the mass difference $\Delta m$
at nuclear matter density $\rho_0$,
for different values of the cutoff mass $\Lambda_D$. First of all, one sees that the net
effect of the in-medium mass change of the $D$ mesons gives a negative shift for
the $J/\Psi$ mass. The total shift ranges $16$ to $24$~MeV at normal nuclear matter
density. The results show in addition that the $D^*D^*$ loop gives the largest
contribution of the three. Also, this contribution is rather insensitive to
the cutoff mass values used in the form factors. A negative self-energy means that the
nuclear mean field provides attraction to $J/\Psi$. The important question is of
course whether such an attraction is enough to bind $J/\Psi$ to a large nucleus.
A partial answer can be obtained as follows. One knows~\cite{Schiff} that
for an attractive spherical well of radius $R$ and depth $V_0$, the condition for
the existence of a nonrelativistic $s$-wave bound state of a particle of mass $m$ is
\begin{equation}
V_0 > \frac{\pi^2 \hbar^2}{8 m R^2}.
\label{bound}
\end{equation}
Using for $m = m^*_{J/\Psi}$ and $R = 5$~fm (radius of a medium-size nucleus),
one obtains $V_0 > 1$~MeV. Therefore, the prospects of capturing a $J/\Psi$ if
produced almost at rest in a nucleus are quite favorable.

Next, we assess the impact of using a gauged Lagrangian for the $DD$ loop on
$m^*_{J/\Psi}$ and $\Delta m$. The results are shown in Table~\ref{tab:xi1}.
The contribution of the $DD$ loop to $\Delta m$ is still much smaller than the
$DD^*$ and $D^*D^*$ contributions, but of opposite sign. The net $J/\Psi$
mass shift is still sizable, varying from $13$~MeV to $18.5$~MeV as the cutoff
is varied from $1000$~MeV to $3000$~MeV. The small, positive
value of the $DD$ loop contribution is in agreement with the result of
Ref.~\cite{Ko:2000jx}.

\begin{table}[htb]
\begin{tabular}{cc|ccc|c}
\hline
$\;\Lambda_D\;$  & $\;m^*_{J/\Psi}\;$  & $\;DD\;$
& $\;DD^*\;$ & $\;D^*D^*\;$ & $\;\Delta m\;$ \\
\hline
                         $1000$ & $3084$ &  $+1$   & $-2$   & $-12$   & $-13$ \\
                         $1500$ & $3082$ &  $+1$   & $-2.5$ & $-12.5$ & $-14$ \\
                         $2000$ & $3080$ &  $+1$   & $-3$   & $-14$   & $-16$ \\
                         $3000$ & $3078$ &  $+0.5$ & $-5.5$ & $-13.5$ & $-18.5$ \\
\hline
\end{tabular}
\caption{Same quantities as in Table~\ref{tab:xi0}, but using the gauged
Lagrangian -- $\xi = 1$ in Eq.~(\ref{KDD}).
}
\label{tab:xi1}
\end{table}

\begin{figure}[htb]
\includegraphics[height=95mm,angle=-90]{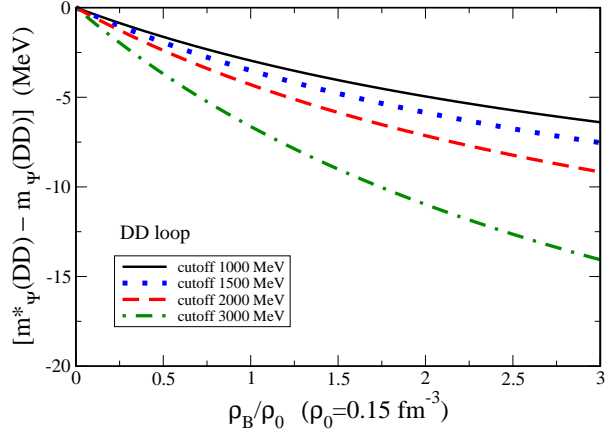}
\caption{Contribution from the $DD$ loop to the difference of the
in-medium and vacuum $J/\Psi$ masses $\Delta m$ as a function of
nuclear matter density for different values of the cutoff mass
$\Lambda_D$.}
\label{fig:DD}
\end{figure}

\begin{figure}[htb]
\includegraphics[height=95mm,angle=-90]{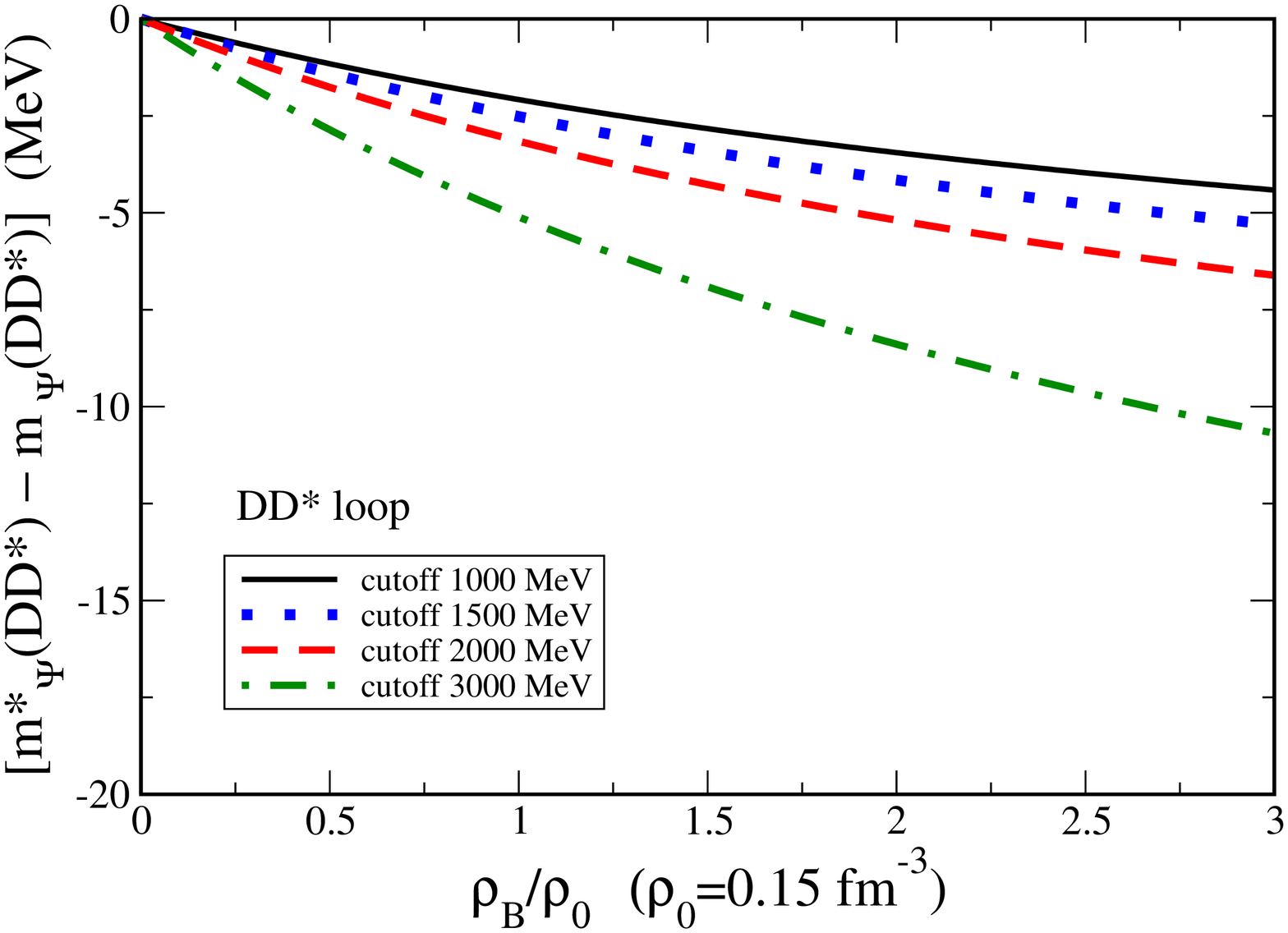}
\caption{Contribution from the $DD^*$ loop. See also caption of Fig.~\ref{fig:DD}.
}
\label{fig:DDs}
\end{figure}

\begin{figure}[htb]
\includegraphics[height=95mm,angle=-90]{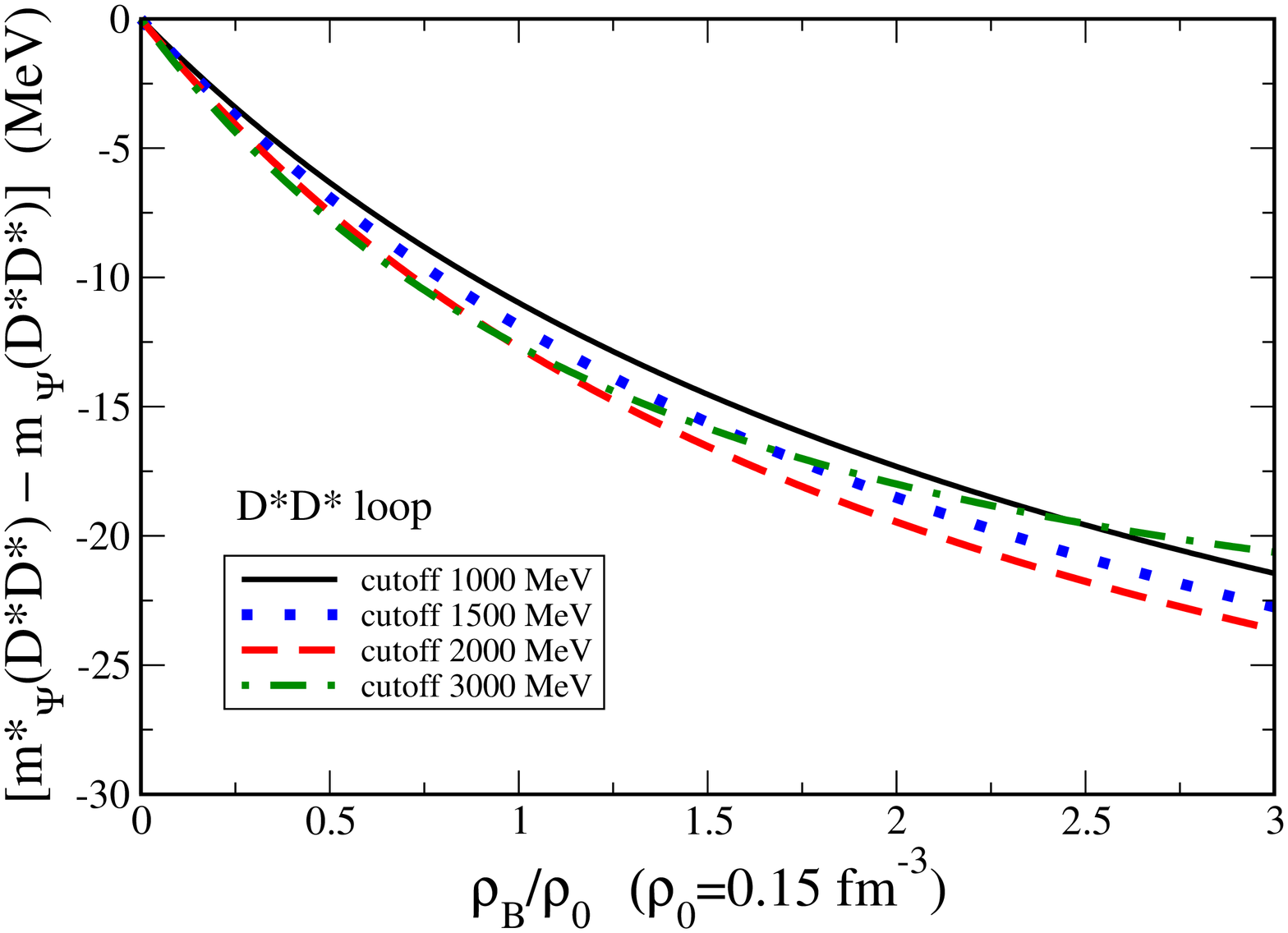}
\caption{Contribution from the $D^*D^*$ loop. See also caption of Fig.~\ref{fig:DD}.
}
\label{fig:DsDs}
\end{figure}

\begin{figure}[htb]
\includegraphics[height=95mm,angle=-90]{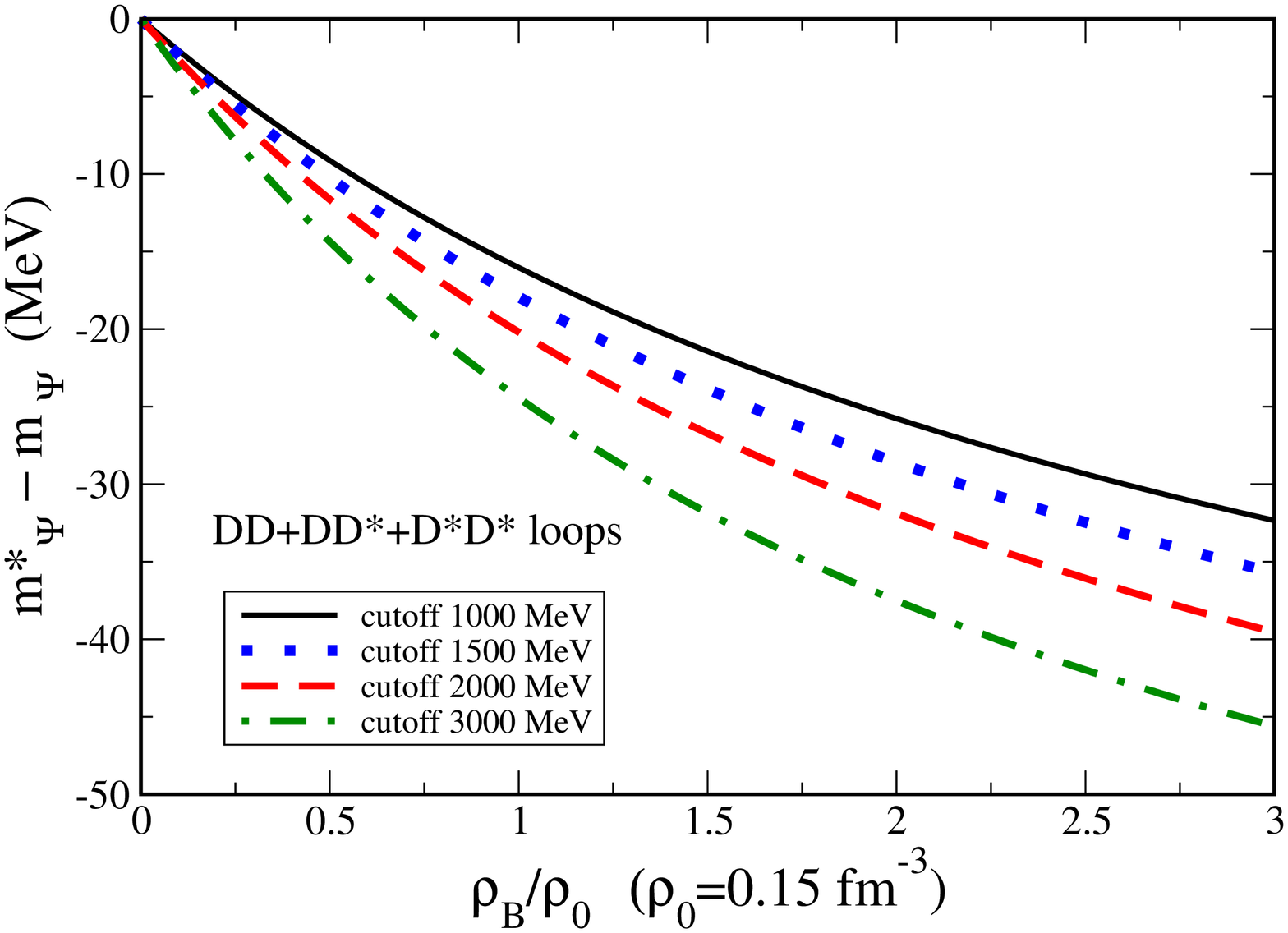}
\caption{The total contributions of the $DD$, $DD^*$ and $D^*D^*$ loops to
the difference of in-medium and vacuum $J/\Psi$ masses
$\Delta m$ as a function of nuclear matter density for different values
of the cutoff mass $\Lambda$.}
\label{fig:total}
\end{figure}

In Figs.~\ref{fig:DD} - \ref{fig:total} we show the separate contributions of
the $DD$, $DD^*$ and $D^*D^*$ loops and their sum to the $J/\Psi$ mass shift.
As the cutoff mass values increase in the form factors, obviously each
loop contribution becomes larger since the integral is divergent, but the
increase is less pronounced for the $D^*D^*$ loop. Since the $D^*D^*$ loop
gives the largest contribution, it is encouraging that this loop contribution
is rather insensitive to the cutoff mass values used.

\section{Conclusions and perspectives}
\label{sec:concl}

We have used an effective Lagrangian approach to evaluate the $D$ and
$D^*$ loop contributions to the mass shift of $J/\Psi$ in cold nuclear
matter. Effects of the medium on the $D$ and $D^*$ are calculated using
the QMC model, in which effective scalar and vector meson mean fields
are coupled to the light $u$ and $d$ quarks in the charmed mesons. There
are no free parameters in this QMC calculation since all quark-meson
coupling constants and bag parameters are fixed by fitting saturation
properties of nuclear matter. The $J/\Psi-D$ coupling constants are taken
as determined from vector meson dominance and the cutoff masses are varied
over a large range of values. The QMC predicts a $62$~MeV mass drop for the
$D$ and $D^*$ mesons at nuclear matter density. This mass drop leads to a
corresponding in-medium $J/\Psi$ mass shift varying between $-16$~MeV and
$-24$~MeV for cutoff masses within the range of $1000$~MeV and $3000$~MeV.
Such a mass shift is large enough to bind a $J/\Psi$ to a nucleus for a
$J/\Psi$ produced at low momentum in the rest frame of the nucleus.

Although the conclusions of the present calculation are very promising
towards the possibility of binding $J/\Psi$ in a nucleus, some issues
clearly require further investigation. Amongst the most important ones
are the calculation of effective $J/\Psi$ potentials for finite
nuclei~\cite{plan} and their momentum dependence, and the inclusion of
$D$ and $D^*$ widths. Recent calculations~\cite{tolos} of in-medium
$D$ and $D^*$ widths based on meson-exchange models have obtained somewhat
contradictory results and further study is required. As emphasized in
Refs.~\cite{DN} the lack of experimental information on the free-space
interaction of $D$ mesons with nucleons is a major impediment for
constraining models and the use of symmetry principles and exploration
of the interplay between quark-gluon and baryon-meson degrees of freedom
is essential in this respect. Still another issue
is the dissociation of $J/\Psi$ in matter by collisions with nucleons and
light mesons. This subject has been studied vigorously in the last years
using different approaches, like meson exchange~\cite{diss-mex} and
quark models~\cite{diss-qm}, QCD sum rules~\cite{Navarra:2001jy}, and
the NJL model~\cite{Bourque:2008es}.
Finally, we stress the need for a deeper understanding of the role
played by color van der Waals forces in the $J/\Psi$ mass shift, particularly
in respect with nucleons interacting in a nucleus.

\acknowledgments
GK thanks the Theory Center of the Jefferson Lab for hospitality and
support during a visit when part of this work was done. The work of GK was
partially financed by CNPq and FAPESP (Brazilian agencies).
Notice: Authored by Jefferson Science Associates, LLC under U.S.
DOE Contract No. DE-AC05-06OR23177. The U.S. Government retains a non-exclusive,
paid-up, irrevocable, world-wide license to publish or reproduce this manuscript
for U.S. Government purposes.
GK and KT would like to acknowledge the hospitality of the CSSM,
University of Adelaide, where the
final part of the calculation was carried out.
This work was also supported by the Australian Research Council through the
award of an Australian Laureate Fellowship to AWT.


\end{document}